\documentclass[sigconf]{acmart}
\AtBeginDocument{%
  }

\usepackage[utf8]{inputenc} 
\usepackage[T1]{fontenc}    
\usepackage{hyperref}       
\usepackage{url}            
\usepackage{booktabs}       
\usepackage{amsfonts}       
\usepackage{nicefrac}       
\usepackage{microtype}      
\usepackage{xcolor}         

\usepackage[boxed,ruled,lined]{algorithm2e}
\usepackage{algorithmic}
\usepackage{multirow}
\usepackage{enumitem}
\usepackage{wrapfig}
\usepackage{subfigure}
\usepackage{amsthm}
\usepackage{amsmath}
\usepackage{cleveref}
\usepackage{caption}
\usepackage{threeparttable}
\usepackage{comment}

\usepackage{setspace}
\usepackage{adjustbox}
\usepackage{bbding}
\usepackage{bm}

\usepackage{graphicx}
\usepackage[normalem]{ulem}
\useunder{\uline}{\ul}{}

\usepackage[short]{optidef}

\definecolor{Instruction}{RGB}{237, 84, 102}
\definecolor{Example}{RGB}{81, 140, 192}
\definecolor{Question}{RGB}{0, 0, 0}
\definecolor{Answer}{RGB}{140, 192, 81}

\newcommand{\red}[1]{{\color{red} #1}}
\newcommand{\blue}[1]{{\color{blue} #1}}

\DeclareMathOperator{\E}{\mathbb{E}}

\setcopyright{acmlicensed}
\copyrightyear{2018}
\acmYear{2018}
\acmDOI{XXXXXXX.XXXXXXX}
\acmConference[Conference acronym 'XX]{Make sure to enter the correct
  conference title from your rights confirmation email}{June 03--05,
  2018}{Woodstock, NY}
\acmISBN{978-1-4503-XXXX-X/2018/06}

\begin{document}

\title{Enhancing Long-Term Welfare in Recommender Systems: An Information Revelation Approach}


\author{Xu Zhao}
\authornote{Both authors contributed equally to this research.}
\affiliation{%
  \institution{Gaoling School of Artificial Intelligence\\Renmin University of China}
  \city{Beijing}
  \country{China}
}
\email{zhao_xu@ruc.edu.cn}

\author{Xiaopeng Ye}
\authornotemark[1]
\affiliation{%
  \institution{Gaoling School of Artificial Intelligence\\Renmin University of China}
  \city{Beijing}
  \country{China}
}
\email{xpye@ruc.edu.cn}

\author{Chen Xu}
\affiliation{%
  \institution{Gaoling School of Artificial Intelligence\\Renmin University of China}
  \city{Beijing}
  \country{China}
}
\email{xc_chen@ruc.edu.cn}

\author{Weiran Shen}
\authornote{Corresponding author.}
\affiliation{%
  \institution{Gaoling School of Artificial Intelligence\\Renmin University of China}
  \city{Beijing}
  \country{China}
}
\email{shenweiran@ruc.edu.cn}

\author{Jun Xu}
\affiliation{%
  \institution{Gaoling School of Artificial Intelligence\\Renmin University of China}
  \city{Beijing}
  \country{China}
}
\email{junxu@ruc.edu.cn}








\renewcommand{\shortauthors}{Xu Zhao et al.}

\begin{abstract}
Improving long-term user welfare (e.g., sustained user engagement) has become a central objective in recommender systems (RS). In real-world platforms, the creation behaviors of content creators play a crucial role in shaping long-term welfare beyond short-term recommendation accuracy, making effective steering of creator behavior essential for fostering a healthy RS ecosystem. Existing approaches typically rely on re-ranking algorithms that heuristically adjust item exposure to influence creators' behavior. However, when integrated into recommendation pipelines, such methods often conflict with the short-term objective of improving recommendation accuracy, leading to performance degradation and suboptimal welfare.

The well-established studies in economics provide valuable insights into an alternative approach without relying on recommendation algorithm design: revealing information from an information-rich party (sender) to a less-informed party (receiver) can effectively influence the receiver's beliefs and steer their behavior.
Inspired by this idea, we propose an information-revealing framework, named \textbf{Lo}ng-term Welfare Optimization via Information \textbf{Re}velation (LoRe). In LoRe, we adopt a classical information revelation method (i.e., cheap talk) to map the stakeholders in RS, treating the platform as the sender and creators as the receivers. To address the challenge posed by the unrealistic assumption of traditional economic method, we formulate the process of information revelation as a Markov Decision Process (MDP) and develop a learning algorithm to optimize the information revelation strategy. Extensive experiments on two real-world RS datasets and across different creator modeling approaches demonstrate that our method can effectively outperform existing re-ranking methods and reward-based creator-influencing method in improving long-term user welfare.
\end{abstract}




\received{20 February 2007}
\received[revised]{12 March 2009}
\received[accepted]{5 June 2009}

\maketitle

\section{Introduction} 
With the rapid growth of user-generated content (UGC)~\cite{krumm2008ugc}, modern recommender platforms (e.g., TikTok, YouTube) have evolved into complex ecosystems where users and content creators continuously interact and influence each other~\cite{bobadilla2013rs_servey, ye2025creagent}. In such dynamic environments, platforms are now shifting their focus from solely optimizing short-term objectives, such as maximizing users’ immediate engagement behaviors (e.g., clicks), to fostering long-term user welfare (i.e., sustained user engagement),  aiming to ensure the lasting health and sustainability of the platform ecosystem~\cite{xu2023pmmf, guo2023fairrec}.

To enhance the long-term user welfare of recommender systems (RS), previous studies have shown that the creators' creation behavior (i.e., creating new items) is a key factor~\cite{ye2025creagent, xu2023pmmf, yao2024user} beyond recommendation accuracy (i.e., prediction accuracy of user behavior). This is because creators continuously upload new items based on previous user feedback, thereby shaping the item pool of RS and indirectly influences what users are exposed to in the long run~\cite{ben2018game,ben2019recommendation,boutilier2024recommender}.
Given the importance of creator behavior, most existing works utilize heuristic re-ranking algorithms to steer creators’ content creation. For example, they increase item exposure for smaller creators to enhance creator retention and stimulate content creation~\cite{guo2023fairrec, xu2023pmmf, ye2024bankfair}. However, prior work have shown that such re-ranking strategies, when embedded within recommendation pipelines, often conflict with the short-term objectives of improving accuracy, inevitably trading off recommendation performance and harming users' short-term experience~\cite{ye2024bankfair, xu2023pmmf}. Consequently, how to steer creators’ content creation behavior without relying on recommendation algorithmic design to enhance the long-term welfare remains a significant challenge in RS.

The investigation of behavioral steering among strategic agents has a long history in economics~\cite{milgrom1987informational,dierkens1991information,mishra1998information,crawford2018asymmetric}, offering valuable insights for enhancing the long-term welfare of RS.
In economics studies, a common approach is to leverage \textbf{information revelation} under the situation of information asymmetry (i.e., agents possess unequal amounts of information) to influence agents’ behavior. 
Such a strategy focuses on designing an information revelation strategy that determines what information an information-rich party (sender) should disclose to a less-informed party (receiver).
By revealing certain information unavailable to the receiver, the receiver can update the beliefs and adjust the action accordingly, resulting in observable behavioral changes.
The revealed information in such situation can be generally interpreted as a suggestion that guides the receiver's action.

Inspired by this idea, we can steer creators’ behavior to enhance long-term welfare by revealing information, viewing the platform as the sender and creators as the receivers. 
Such a strategy is feasible and practical in real-world platforms, because information asymmetry naturally exists between the platform and creators in RS~\cite{prasad2023content,boutilier2024recommender,ye2025creagent}. Similar but simpler idea have been adopted in industrial practice, offering real-world scenarios and potential applications for more advanced information revelation approaches. For instance, platforms like TikTok and YouTube provide creators with suggestions to produce trending content.

However, directly applying information revelation to RS is challenging. Traditional economic approaches~\cite{yao2024user,prasad2023content,boutilier2024recommender} typically assume that receivers are completely rational, i.e., they always select the optimal strategy to maximize their utility. Under this assumption, receivers' behavior can be computed explicitly, enabling the sender to optimally send information to maximize its own utility. In contrast, real-world creators are boundedly rational~\cite{simon1990bounded, kahneman2013prospect}, because their decision-making is constrained by the limited cognitive capacity, attention, and information-processing ability of humans. Consequently, it is difficult to calculate creators’ behavior in real-world precisely, preventing the platform from directly computing the optimal information to reveal.

To address this challenge, we propose an information revelation framework tailored for RS, named \textbf{Lo}ng-term Welfare Optimization via Information \textbf{Re}velation (LoRe). Specifically, we map the key components of classical information revelation method in economics studies (i.e., cheap talk model~\cite{farrell1996cheap}) onto the corresponding stakeholders in RS. Based on this mapping, we formulate the platform’s information revelation process as a Markov Decision Process (MDP)~\cite{puterman1990markov} and develop a learning algorithm to optimize the platform's information revelation strategy.

Unlike previous works that rely on heuristic exposure allocation in the re-ranking stage to steer creators’ behavior~\cite{guo2023fairrec}, LoRe serves as an independent module for the platform and does not depend on any specific recommendation algorithm.
Experiments on two real-world recommendation datasets demonstrate that LoRe outperforms both creator-fair re-ranking methods and reward-based creator-influencing method in improving long-term user welfare (i.e., sustained user clicks).
Moreover, since LoRe operates independently, it can be seamlessly integrated with existing ranking or re-ranking algorithms to further enhance long-term welfare.

In our experiments, to more faithfully simulate the real-world RS environments, we adopt the multi-stakeholder simulation framework proposed by~\citet{ye2025creagent}, which leverages LLM-based agents to simulate the dynamic behaviors of users and creators. Specifically, these agents simulate user click-through behavior and creator content generation. This approach is grounded in recent research which demonstrate that agents empowered by LLMs can effectively emulate user preferences and the creation behavior of bounded rational creators~\cite{wang2024macrec, zhang2024generative, wang2025user, ye2025creagent}.

We summarize our contributions as follows:

(1) We highlight that influencing creators' creation behavior through information revelation provides an effective approach to enhancing long-term user welfare beyond relying on recommendation algorithm design.

(2) Inspired by economic studies, we propose a novel information revelation method, LoRe, which models the revelation process as a MDP and learns to optimize the information revelation strategy in environments where boundedly rational agents interact.

(3) Extensive experiments on real-world datasets demonstrate that our method can effectively improve the long-term user welfare of RS by steering creators' creation behavior.
\section{Related Work}

\textbf{Multi-stakeholder RS.}
Multi-stakeholder RS have received significant attention in recent years in both academic research and industrial practice~\cite{abdollahpouri2020multistakeholder_servey, chen2025creator_side_RS}.
Unlike traditional RS, which primarily focuses on providing the most suitable items based on each user’s request~\cite{he2017ncf,rendle2012bpr}, multi-stakeholder RS also takes into account the interests of platforms and creators, aligning with the long-term goals of many multi-stakeholder real-world platforms (e.g., TikTok, YouTube).
Most existing studies on multi-stakeholder RS can be broadly categorized into two directions: (1) designing better RS algorithms that aim to achieve multi-stakeholder objectives (e.g., provider fairness~\cite{ye2024bankfair, xu2025fairdiverse}, platform profit~\cite{pei2019value-aware,azaria2013movieRS4profit}) while maintaining recommendation accuracy; (2) incorporating creators and platforms into the RS environment to enable more comprehensive evaluation of multi-stakeholder RS. For example, some studies focused on developing re-ranking algorithms that allocate exposure resources fairly among different providers to achieve different provider fairness objectives (e.g., Gini coefficient~\cite{do2022optimizinggini}, Max-Min Fairness~\cite{ye2024bankfair,xu2023pmmf}). 
\citet{ye2025creagent} introduces an LLM-based creator simulator, CreAgent, aiming to achieve a more realistic evaluation of multi-stakeholder RS.
However, existing works mainly focus on developing and evaluating RS algorithms, and do not consider another crucial side, i.e., steering creators via information revelation to indirectly optimize the long-term welfare of RS.

\textbf{Information asymmetry in RS and information revelation.}
A relevant topic to our work is the literature considering the information asymmetry in RS~\cite{prasad2023content,yao2023rethinking,yao2024user,boutilier2024recommender,ye2025creagent}. \citet{boutilier2024recommender,ye2025creagent} highlight the significant information asymmetry in RS, but they do not proposed a method to address such a problem. \citet{prasad2023content} propose a prompting policy that promises to provide creators a certain level of user clicks to influence the their creation behavior. \citet{yao2023rethinking,yao2024user} suggest influencing content creation by adjusting reward weights to encourage different types of content. However, these methods inevitably harm creators' realized rewards and may introduce unfairness, as they directly alter the rewards that creators would otherwise receive.
The literature on information revelation under information asymmetry in economics is also related to our work, where a prominent approach is cheap talk~\cite{farrell1996cheap,teacy2006agent,chakraborty2010persuasion,ivanov2015dynamic}. In cheap talk, an informed party (sender) designs an information revelation strategy to reveal the information to a less informed party (receiver). However, because of their strong assumption of completely rational receivers, these models cannot be directly applied to real-world RS scenarios where creators typically exhibit bounded rationality.

\textbf{LLM-based simulator in RS.}
In recent years, given the superior semantic understanding and reasoning capabilities of LLMs, LLM-based simulators for recommender systems have attracted increasing attention and experienced rapid development. They are now widely regarded as a reliable technique for modeling user behavior in recommender systems~\cite{ye2025creagent, ma2025pub,zhu2025llm,zhang2025llm}. Initially, research~\cite{ma2025pub,zhu2025llm,zhang2025llm} focused on simulating user behavior in RS, such as clicks, to obtain real-world user feedback for evaluating recommendation algorithms. Most recently, the field has progressed toward multi-stakeholder and ecosystem simulation, recognizing that RS operate within complex environments involving users, creators and platforms~\cite{wang2024macrec,ye2025creagent,coppolillo2025engagement}. These studies demonstrated that LLM-based RS simulators can replicate the behaviors of both creators and users in real-world, providing a reliable framework to evaluate the performance of algorithms developed for such systems.
\section{Preliminaries}

\subsection{Recommender Systems}\label{sec:RS}

\textbf{Stakeholders in RS.} Considering a real-world content recommendation platform (e.g., YouTube, TikTok). The platform needs to recommend items (e.g., videos, articles, etc.) produced by content creators to users. Let $\mathcal{U} = \{u_i\}_{i=1}^{|\mathcal{U}|}$, $\mathcal{C}=\{c_i\}_{i=1}^{|\mathcal{C}|}$, $\mathcal{G}=\{g_i\}_{i=1}^{|\mathcal{G}|}$ represent the set of users, creators, and item genres that creators can produce. Besides, let $\mathcal{I}$ denote the current corpus of items that can be recommended to users.

\textbf{Interactions between different stakeholders.} We model the interaction between the platform, creators, and users as a dynamic process with total $T$ interaction rounds. At the beginning of each round $t\in \{1,2,\dots, T\}$, each creator $c_j$ will produce an item with the genre in $\mathcal{G}$. The items produced by creators in each round will be added to $\mathcal{I}$. Then, some users will arrive and access the platform. For each accessing user, the platform needs to select $k$ items from $\mathcal{I}$ for recommendation. Each user can choose to click or skip the items recommended to him. After that, the platform records the interaction between users and items. Specifically, we use $\bm{Y}$ to represent the click matrix, where $\bm{Y}_{i,j,t}=1$ if item $i\in \mathcal{I}$ is clicked by user $u_j$ in round $t$ and 0 otherwise. In addition, we use $\mathcal{H}_{c_j}=\{(i,k_i)\mid i\in\mathcal{I}\}$ to denote historical information of items created by creator $c_j$, where $(i,k_i)\in \mathcal{H}_{c_j}$ means that the item $i\in \mathcal{I}$ produced by creator $c_j$ has been clicked $k_i$ times. For easy notation, let $\mathcal{H}=\{\mathcal{H}_{c_1},\mathcal{H}_{c_2},\dots,\mathcal{H}_{c_{|\mathcal{C}|}}\}$ denote the 
set of all creators' historical information.

\textbf{Information asymmetry in RS.} In real-world RS, the platform has the historical information $\mathcal{H}$ of all creators and the interaction information $\bm{Y}$ of all users and items. In contrast, each creator $c_j$ only has its own historical information $\mathcal{H}_{c_j}$. Obviously, there is a significant information asymmetry between platform and creators.

\begin{figure*}[ht]
    \centering
    \includegraphics[width=0.90\linewidth]{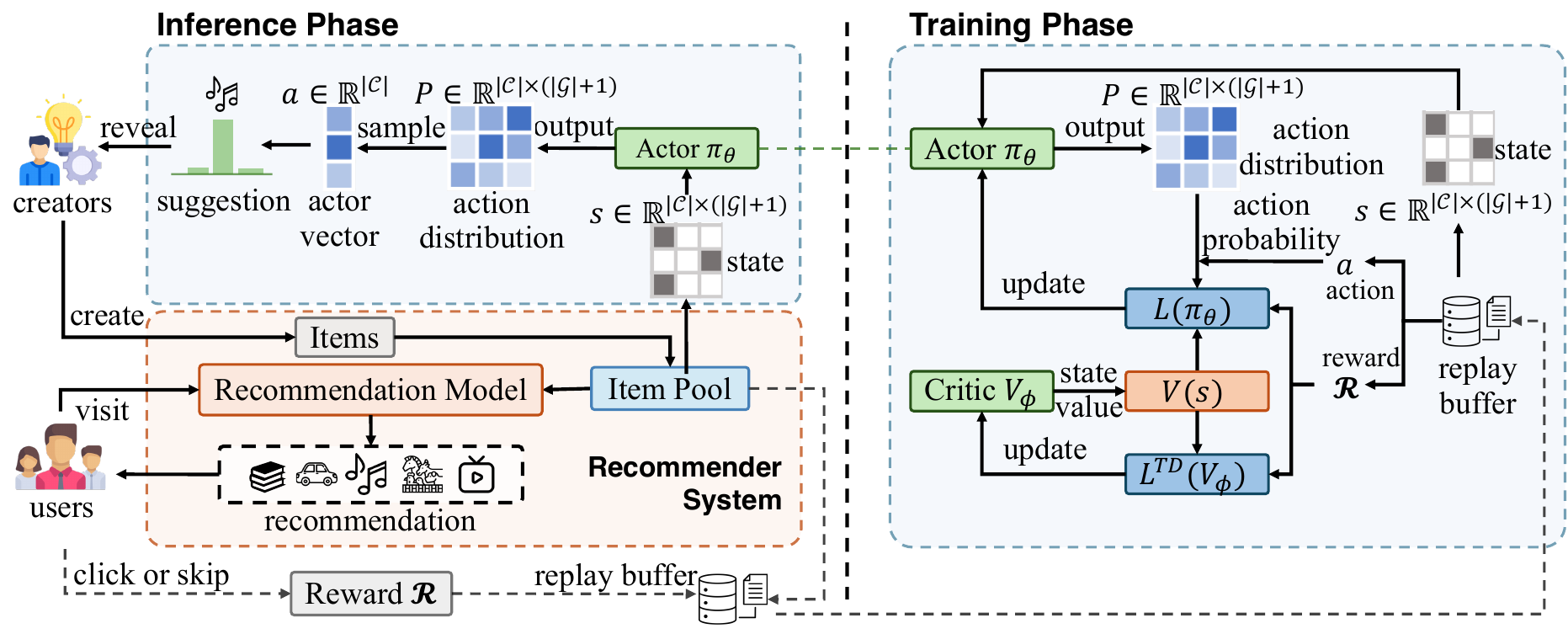}
    \caption{The overall framework of our method LoRe, which consists two phases: Inference and Training. During the inference phase, in each round, the platform sends suggestions to creators, who then produce content, after which users decide whether to click on the recommended items. The interaction data generated in each round are collected by the platform. In the training phase, the platform uses the data collected during inference to train the actor and critic networks.}
    \label{fig: framework}
    \vspace{-0.3cm}
\end{figure*}

\subsection{Cheap Talk in Recommender Systems}
\textbf{Cheap talk.}
The standard cheap talk model considers an idealized setting with a sender and a receiver. Both the receiver's utility function $\mu_r (a_r,w)$ and the sender's utility function $\mu_s(a_r,w)$ depend on receiver's action $a_r\in \mathcal{A}_r$ and state $w\in \mathcal{W}$.
In this model, only the sender exactly knows the true state $w$, whereas the receiver does not. Thus, the sender possesses more information than the receiver, giving rise to information asymmetry between sender and receiver.

When the receiver has no additional information, she chooses an optimal action $a_r^*$ to maximize her expected utility based on what she knows, and the sender correspondingly obtains utility $\mu_s(a_r^*,w,b)$. However, the sender can influence the receiver’s action by revealing information (message). Specifically, we use $M$ to denote the set of possible messages the sender can send to receiver. The specific form of a message is immaterial, as long as it is commonly understood and interpretable by both parties. Intuitively, a specific message $m\in M$ can be viewed as a suggestion of the form: \textbf{"Based on the available information, I recommend that you take action $\hat{a}_r$"}.
Formally, we use $\pi:\mathcal{W}\times M \mapsto [0,1]$ to denote the sender's information revelation strategy, where $\pi(m\mid w)$ represents the probability of sending message $m\in M$ when the real state is $w$. Upon receiving a message, the receiver updates her beliefs and chooses an optimal action accordingly. Previous researches~\cite{farrell1996cheap,teacy2006agent,chakraborty2010persuasion,ivanov2015dynamic} shows that, by appropriately designing the information revelation strategy, the sender can improve the utilities of both the sender and the receiver under information asymmetry.

\begin{table}
    \centering
\caption{Correspondence between elements in Cheap Talk and RS. See section~\ref{sec:approach} for undefined notations.}
    \begin{tabular}{c|c}
    \toprule
    Cheap Talk   &  Recommender System \\
         \hline
    sender & platform \\
    receivers & creators\\
    state $w$ & distribution of items $s$\\
    message set $M$  & suggestion set $\mathcal{A}_p$\\
    information revelation strategy $\pi$ & suggestion strategy $\pi_{\theta}$\\
        
        \bottomrule
    \end{tabular}
    \label{tab:correspond}
\end{table}

\textbf{Cheap talk in Recommender Systems.}
Motivated by the advantage of cheap talk under information asymmetry, we adopt this framework to guide how the platform reveals information in RS, where information asymmetry similarly exists between the platform and content creators. Simply put, we allow the platform to send suggestions to creators, such as \textbf{"I suggest that you create an item of genre $g_i\in \mathcal{G}$, as this may attract more clicks or likes"}. We map the key components of RS to their counterparts in the cheap talk model, as summarized in Table~\ref{tab:correspond}. A detailed description of these correspondences is provided below.

We model the platform and the creators as the sender and the receivers, respectively. This modeling choice is motivated by the fact that the platform possesses richer information and can influence creators' creation behavior through information revelation. Moreover, creators' creation behavior shapes the item pool of RS, thereby indirectly determining the content to which users are exposed over time and, consequently, affecting long-term user welfare. Therefore, by steering creators' creation behavior, platforms can indirectly improve long-term user welfare. 

In the cheap talk model, the receiver’s utility depends on the actual state, which is unobserved by the receiver. Analogously, in RS, we represent the state as the distribution of item genres, since this distribution has a substantial impact on clicks, likes, and other performance metrics of creators' items, and is typically unknown to individual creators. For example, if an increasing number of creators produce items of the same genre, competition within that genre intensifies, reducing the traffic received by each creator. We denote the set of possible states by $\mathcal{S}$.

As the platform is allowed to send suggestions to creators, the messages in the cheap talk model correspond to the platform’s suggestions in RS. We use $\mathcal{A}_p=\{a_i\}_{i=0}^{|\mathcal{G}|}$ to denote the set of platform's possible suggestions, where $a_0$ represents sending no suggestion, and $a_i\ (i\neq0)$ represents the suggestion: "I recommend that you create item with genre $g_j$". 

Corresponding to the sender’s information revelation strategy in the cheap talk model, in RS, we use $\pi_{\theta}:\mathcal{S}\times \mathcal{A}_p \mapsto [0,1]$ to represent the platform's suggestion strategy. Specifically, $\pi_{\theta}(a_i\mid s)$ is the probability that the platform sends suggestion $a_i$ when the state is $s\in \mathcal{S}$.

In the traditional cheap talk model, the sender can compute an optimal information revelation strategy to maximize his own utility because the receiver is assumed to be completely rational, and thus the receiver's behavior in response to any message can be precisely inferred. In contrast, in RS, creators are not completely rational and their creation behavior after receiving platform's suggestions is generally unknown to the platform. As a result, the platform cannot directly compute the optimal suggestion strategy. Consequently, the theoretical results of the classic cheap talk model cannot be directly applied to RS, making the optimization of the platform’s suggestion strategy a challenging problem.
\section{Our Approach: LoRe}\label{sec:approach}
In this section, to address the challenge by applying the framework of cheap talk into RS and optimizing the platform's suggestion strategy, we propose a novel method, LoRe. The overall algorithm workflow of LoRe is shown in Figure~\ref{fig: framework} and Algorithm~\ref{alg:optimization}. We first formulate the dynamic interaction process between the platform and creators under information asymmetry as a Markov decision process (MDP), which is described in Section~\ref{sec:MDP}. We then design a reinforcement learning algorithm to optimize the platform's suggestion strategy in this dynamic environment, as presented in Section~\ref{sec:RLA}.

\subsection{Interactions Between Platform and Creators}\label{sec:MDP}
We now formulate the dynamic interaction between the platform and creators in each round as follows. At the beginning of round $t$, the platform observes the state $s^t\in \mathcal{S}$, which represents the distribution of item genres. The platform then sends each creator a suggestion from suggestions set $\mathcal{A}_p$. Upon receiving a suggestion, a creator can choose to produce an item with any possible genre $g_i\in \mathcal{G}$. We do not impose any assumption of creator's creation behavior, which indicates that any creator's behavior model can be embedded in our method. After content creation and recommendation, creator $c_i$ obtains the number of clicks of his item as utility, denoted by $\mu_{c_i}^t$.
Then, the cumulative utility of creator $c_i$ over $T$ rounds is given by:
\begin{gather*}
    \mu_{c_i} = \sum_{t=1}^T \mu_{c_i}^t.
\end{gather*}
The platform's objective is to maximize the long-term user welfare, measured by the total number of user clicks, through sending appropriate suggestions. With a slight abuse of notation, we use $\bm{a}^t = (a_1^t,a_2^t,\dots,a_{|\mathcal{C}|}^t)$ to denote the platform's suggestion strategy in round $t$, where $a_i^t\in \mathcal{A}_p$ is the suggestion sent to creator $c_i$. Then, the platform's problem can be formulated as follows:
\begin{gather*}
    \max_{(\bm{a}^1,\bm{a}^2,\dots,\bm{a}^T)} \sum_{c\in \mathcal{C}}\mu_c.
\end{gather*}


This interaction can be abstracted as follows. Given a state, the platform sends suggestions to creators, after which the system transitions to a new state as a result of creators’ creation behavior, and the platform receives a reward equal to the total number of user clicks. Moreover, because the state is defined by the distribution of item genres, the state transition can be reasonably approximated as depending only on the current state and the platform’s actions. Consequently, the dynamic interaction between the platform and creators can be naturally modeled as a Markov decision process. We formulate this MDP between the interaction between platform and creators as a tuple $M=(\mathcal{S},\mathcal{A},\mathcal{T},\mathcal{R}, \gamma)$, where:\\
\textbf{$\bullet$ State $\mathcal{S}$:} $\mathcal{S}$ is the set of states. Each state is denoted by
\begin{align*}
    \bm{s}^t=(e_1^t,e_2^t,\dots,e_{|\mathcal{C}|}^t)\in \mathcal{S},
\end{align*}
which is the distribution of item genres in round $t-1$, where $e_i^t$ denotes the genre of item created by creator $c_i$ in round $t-1$.\\
\textbf{$\bullet$ Action $\mathcal{A}$:} $\mathcal{A}$ is the set of the platform's actions. Platform's each action $\bm{a}^t\in \mathcal{A}$ is a vector $(a_1^t,a_2^t,\dots,a_{|\mathcal{C}|}^t)$, where $a_i^t\in \mathcal{A}_p$ represents the suggestion sent to creator $c_i$ in round $t$.\\
\textbf{$\bullet$ Transition function $\mathcal{T}$:} The transition probability function is defined as $\mathcal{T}:\mathcal{S} \times \mathcal{A}\times \mathcal{S}\mapsto[0,1]$, where $\mathcal{T}(\bm{s}^{t+1}|\bm{s}^{t},\bm{a}^t)$ denotes the probability that the state at round $t+1$ is $\bm{s}^{t+1}$ given that the the state at round $t$ is $\bm{s}^{t}$ and the platform takes action $\bm{a}^t$.\\
\textbf{$\bullet$ Reward function $\mathcal{R}$:} The reward function is defined as $\mathcal{R}:\mathcal{S} \times \mathcal{A} \times \mathcal{S} \mapsto \mathbb{R}$, where $\mathcal{R}(\bm{s}^{t},\bm{a}^t,\bm{s}^{t+1})$ denote the total user clicks (the measure of welfare) in round $t$, given that state at round $t$ is $\bm{s}^t$, the platform takes action $\bm{a}^t$, and the state transitions to $\bm{s}^{t+1}$ in round $t+1$.\\
\textbf{$\bullet$ Discount rate $\gamma$:} $\gamma\in [0,1]$ is the discount factor that quantifies the present value of future rewards.

The platform’s suggestion strategy is redefined as:
\begin{gather*}
    \pi_{\theta}:\mathcal{S}\times\mathcal{A}\mapsto[0,1],
    \label{eq:strategy}
\end{gather*}
where $\pi_{\theta}(\bm{a}^t|\bm{s}^{t})$ is the probability of taking action $\bm{a}^t$ when the state at round $t$ is $\bm{s}^{t}$. Then, the expected discounted user clicks under strategy $\pi_{\theta}$ can be formulated as:
\begin{gather}
    J(\pi_{\theta}) = \E_{\tau\sim\pi_{\theta}}\left[\sum_{t=0}^{\infty}\gamma^{t}\mathcal{R}(\bm{s}^{t},\bm{a}^t,\bm{s}^{t+1})\right],
    \label{eq:welfare}
\end{gather}
where $\tau$ is a trajectory, i.e., $\tau=(\bm{s}^0,\bm{a}^0,\bm{s}^1,\bm{a}^1,\dots)$ and $\tau\sim\pi_{\theta}$ is a shorthand indicating that the distribution over trajectories depends on $\pi_{\theta}:\bm{a}^t\sim\pi_{\theta}(\cdot|\bm{s}^{t})$, $\bm{s}^{t+1}\sim \mathcal{T}(\cdot|\bm{s}^{t},\bm{a}^t)$. Furthermore, we respectively use $V^{\pi_{\theta}}(\bm{s}^t)$, $Q^{\pi_{\theta}}(\bm{s}^{t},\bm{a}^t)$ and $A^{\pi_{\theta}}(\bm{s}^t,\bm{a}^t)$ to denote the state value function, state-action value function and the advantage function, where $V^{\pi_{\theta}}(\bm{s}^t)$ is the value of state $\bm{s}^t$, $Q^{\pi_{\theta}}(\bm{s}^{t},\bm{a}^t)$ is the value of taking action $\bm{a}^t$ under state $\bm{s}^t$ and $A^{\pi_{\theta}}(\bm{s}^t,\bm{a}^t)=Q^{\pi_{\theta}}(\bm{s}^t,\bm{a}^t)-V^{\pi_{\theta}}(\bm{a}^t)$\footnote{In practice, the advantage function $A^{\pi_{\theta}}(\bm{s}^t,\bm{a}^t)$ is calculated by the generalized advantage estimation~\cite{schulman2015high}.}. $V^{\pi_{\theta}}(s^t)$ and $Q^{\pi_{\theta}}(\bm{s}^{t},\bm{a}^t)$ are common functions in MDP, and their definitions are as follows:
\begin{gather*}
    V^{\pi_{\theta}}(\bm{s}^{t})=\E_{(\bm{a}^t,\bm{s}^{t+1},\dots)\sim\pi_{\theta}}\left[\sum_{l=0}^{\infty}\gamma^l\mathcal{R}(\bm{s}^{t+l},\bm{a}^{t+l},\bm{s}^{t+l+1})\right],\\
    Q^{\pi_{\theta}}(\bm{s}^{t},\bm{a}^t) = \E_{(\bm{s}^{t+1},\bm{a}^{t+1},\dots)\sim\pi_{\theta}}\left[\sum_{l=0}^{\infty}\gamma^l\mathcal{R}(\bm{s}^{t+l},\bm{a}^{t+l},\bm{s}^{t+l+1})\right].
\end{gather*}

\subsection{Reinforcement Learning Algorithm for Platform's Suggestion Strategy}\label{sec:RLA}
In this section, we aim to optimize the long-term welfare in Eq(\ref{eq:welfare}) under the dynamic user-creator-platform interaction environment by learning an suggestion strategy $\pi_{\theta}$. To achieve this, we design an actor-critic based RL algorithm. The overall workflow of our algorithm is shown in Algorithm~\ref{alg:optimization}, which contains two phases: the inference phase (Line 4-13) and the training phase (Line 14-19). 

\begin{algorithm}[t]
        \caption{The overall workflow of our method LoRe.}
    	\label{alg:optimization}
    	\begin{algorithmic}[1]
    	\REQUIRE Discount rate $\gamma$, clip hyperparameter $\epsilon$, learning rate $lr$, total rounds $T$, training epochs $M$, cumulative data rounds $N$, reply buffer $\mathcal{B}=\{(\bm{s}^t,\bm{a}^t,\bm{s}^{t+1},\mathcal{R})\}$.

\ENSURE The policy loss $L(\pi_{\theta})$ decreases by less than 1\% over 10 consecutive rounds.
\STATE Initialize state $\bm{s}^1$ from log data.
\FOR{$t=1$ to $T/N$}
\STATE Clear reply buffer: $\mathcal{B}=\{\}$.
\STATE $// ~~\texttt{\red{Inference Phase}}$.
\FOR{$r=1$ to $N$}
\STATE Get state $\bm{s}^t$.
\STATE Sample action $\bm{a}^t\sim\pi_{\theta}(\bm{s}^t)$. $// ~~\texttt{Platform sends suggestions}$.
\STATE $// ~~\texttt{Creators produce items}$.
\STATE The next state transitions to $\bm{s}^{t+1}$.
\STATE $// ~~\texttt{Users interact with items (click or skip)}$.
\STATE Get the total users' clicks $\mathcal{R}$.
\STATE Add $(\bm{s}^t,\bm{a}^t,\bm{s}^{t+1},\mathcal{R})$ to reply buffer $\mathcal{B}$.
\ENDFOR
\STATE $// ~~\texttt{\blue{Training Phase}}$.
\FOR{$i=1$ to $M$}
\STATE Calculate actor loss $L$ and critic loss $L^{TD}$ based on $\mathcal{B}$ according to Eq.~(\ref{eq:clip objective}) and Eq.~(\ref{eq:TD}).
\STATE Update actor network using the gradient $\nabla_{\pi_{\theta}}L$.
\STATE Update critic network using the gradient $\nabla_{V_{\phi}}L^{TD}$.
\ENDFOR
\ENDFOR
\end{algorithmic}
\end{algorithm}

\subsubsection{Inference Phase}
At the beginning, the platform gets the current item genres' distribution, which constitute the state $\bm{s}^t$. Then, the actor network $\pi_{\theta}$, which is parameterized by $\theta$, takes $\bm{s}^t$ as input and outputs the action distribution:
\begin{equation*}
    \bm{P}^t=\pi_{\theta}(\bm{s}^t).
    \label{eq:P}
\end{equation*}
where the input $\bm{s}^t$ and output $\bm{P}^t$ are illustrated as follows.

\textbf{Input.} As shown in Figure~\ref{fig: framework}, we represent the state $\bm{s}^t\in \mathbb{R}^{|\mathcal{C}|\times (|\mathcal{G}|+1)}$ as a one-hot matrix, where each row of the matrix is the one-hot vector of the item created by corresponding creator. The length of the one-hot vector is $|\mathcal{G}|+1$, because there are cases where the creator dose not create item.

\textbf{Output.} In the classic actor network, the actor usually outputs a distribution over the action space $\mathcal{A}$. However, the number of possible actions is $|\mathcal{A}|=(|\mathcal{G}|+1)^{|\mathcal{C}|}$, which is too large to handle. To address this problem, we design the actor to output the distribution over suggestion set $\mathcal{A}_p$ rather than the distribution over action set $\mathcal{A}$. That is, instead of outputting a joint distribution over all platform's possible actions, we output an independent distribution over the suggested actions for each creator. Specifically, the output of actor is $\bm{P}=\pi_{\theta}(\bm{s}^t)\in\mathbb{R}^{|\mathcal{C}|\times(\mathcal{G}|+1)}$, where each row of $\bm{P}$ is a distribution over the suggestion set for corresponding creator. The suggestions sent to creators are sampled from the corresponding row of $\bm{P}$, which constitute the platform's action $\bm{a}^t\in \mathbb{R}^{|\mathcal{C}|}$.

After sending suggestions to creators, creators produce items and the genres of these new items constitute the next state $\bm{s}^{t+1}$. After that, platform recommends items to each user and get the total users' clicks $\mathcal{R}$ as reward. Finally, the data $b^t=(\bm{s}^t,\bm{a}^t,\bm{s}^{t+1},\mathcal{R})$ is then added to the reply buffer $\mathcal{B}$ which is used to train the actor and critic networks in the training phase.

\subsubsection{Training Phase}
We use the collected reply buffer $\mathcal{B}$ to train the actor and critic networks. We denote the critic network by $V_{\phi}(\bm{s})$, which is parameterized by $\phi$. The critic network takes state $\bm{s}^t$ as input and outputs the estimated state value of $\bm{s}^t$, which is used to approximate the state value function $V^{\pi_{\theta}}(\bm{s})$ and help optimize the actor network.

\textbf{Actor update.} To update the actor, we use the clipped surrogate objective~\cite{schulman2017proximal} as the loss function:
\begin{gather}
    L(\pi_{\theta})= - \sum_{b\in \mathcal{B}}\min\left(\pi_{\theta}'(\bm{a}|\bm{s}),\text{clip}\left(\pi_{\theta}'(\bm{a}|\bm{s}),1-\epsilon, 1+\epsilon\right)\right)A^{\pi_{\theta}}(\bm{a},\bm{s}),
    \label{eq:clip objective}
\end{gather}
where $\pi_{\theta}'(\bm{a}|\bm{s})=\frac{\pi_\theta(\bm{a}|\bm{s})}{\pi_{\theta_{old}}(\bm{a}|\bm{s})}$, $\text{clip}(x,l,r)=\max(\min(x,r),l)$ and $\epsilon$ is the clip hyperparameter. $\pi_{\theta_{old}}$ is the actor network after the last update and remains unchanged when the current actor network $\pi_{\theta}$ is updated. 

\textbf{Critic update.} To update the critic network, we use the MSE of the temporal-difference error (TD-error) as the loss function:
\begin{gather}
    L^{TD}(V_{\phi}) =  \sum_{b\in \mathcal{B}} (\mathcal{R}_t+\gamma V_{\phi}(\bm{s}^{t+1})-V_{\phi}(\bm{s}^t))^2.
    \label{eq:TD}
\end{gather}

We assess convergence based on the policy loss $L(\pi_{\theta})$. Specifically, the algorithm is considered converged when the policy loss remains approximately stable, operationalized as a decrease of less than 1\% over 10 consecutive episodes.
\begin{figure}
\includegraphics[width=0.45\textwidth]{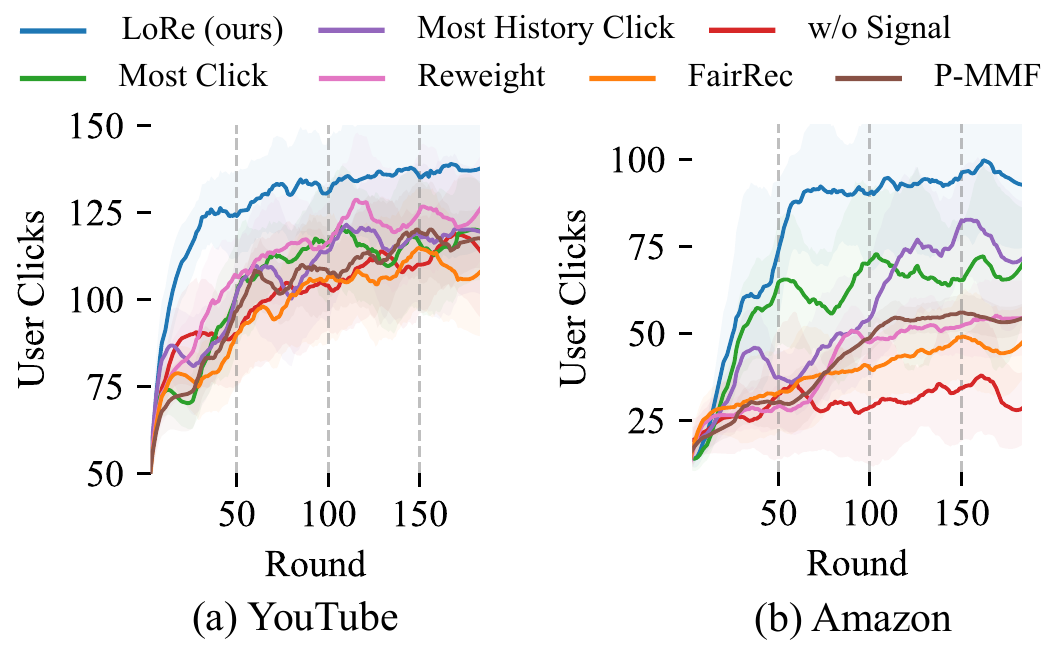}
	\caption{Comparison of user clicks in RS between our method and other baselines on two different datasets. }
	\label{fig:RQ1_main}
\end{figure}
\section{Experiments}\label{sec:exp}
In this section, we conduct extensive experiments to evaluate the effectiveness of our method in improving the long-term welfare of RS. Specifically, we aim to address the following research questions::\\
\textbf{RQ1}: Can our method improve the long-term welfare of RS through information revelation?\\
\textbf{RQ2}: What is the inherent insight that our method can enchance the long-term welfare?\\
\textbf{RQ3}: How robust is our method, and how does it perform under various settings?

The source code of LoRe and the datasets used in our experiments are publicly available at \textcolor{blue}{\url{https://anonymous.4open.science/r/LoRe-E5F0}}.

\subsection{Experimental Setups}

\textbf{Real-world dataset.}
We conduct experiments on two large-scale recommendation datasets, both collected from real-world recommendation platforms.
\textbf{YouTube} is a dataset crawled from the world’s leading recommendation platform~\cite{ye2025creagent}. It contains rich information about content creators across diverse categories, along with user comments on their uploaded items. The dataset comprises 4,004 creators, 1.97 million users, and 0.19 million distinct items spanning 14 genres.
\textbf{Amazon Video Games} (Amazon)~\footnote{\url{https://amazon-reviews-2023.github.io/}} is a widely used recommendation dataset. It consists of user-generated reviews, ratings, and product metadata collected from the Amazon e-commerce platform. It contains 2.8M users, 137.2K items, and 1326 creators.

\textbf{Simulation setups.} To better simulate the real-world RS environments, we adopt the multi-stakeholder RS simulation platform proposed by~\citet{ye2025creagent}, which integrates LLM-based agents to simulate the dynamic behaviors of both users and creators. Recent studies have demonstrated that LLMs are capable of accurately mimicking user preferences and the decision-making of bounded-rational creators~\cite{wang2024macrec, zhang2024generative, wang2025user, ye2025creagent}. 
To evaluate the robustness and generalization ability of our method under diverse environments, we additionally conduct experiments under environments where creators are simulated using non-LLM-based models (See Section~\ref{sec:creator_ablation}).
Following previous works~\cite{mladenov2020optimizing,xu2023pmmf,ye2025creagent}, we also consider the active behaviors of creators and users, as well as creator retention. Specifically, each creator $c_i\in\mathcal{C}$ and each user $u_i\in\mathcal{U}$ has probability $p_{c_i}$ and $p_{u_i}$ to be active, respectively. In each round, these probabilities determine whether a creator or user is active. Only active creators create items, and only active users receive recommendations. Moreover, a creator $c_i$ is assumed to leave the platform if no clicks are received in $l$ consecutive rounds of creation.

\begin{table*}[htb]
\centering
\caption{Comparisons of average user clicks, content diversity and active creators of our methods and baselines on YouTube and Amazon datasets. MHC represents the most history click and MC denotes the most click. We categorize baselines into three types: naive suggestion strategy, re-ranking methods, and reward adjust method.}
\resizebox{0.99\textwidth}{!}{
\begin{tabular}{llcccccccc}
\toprule
  &  & \multicolumn{2}{c}{\textbf{User Clicks}} & & \multicolumn{2}{c}{\textbf{Content Diversity}}& & \multicolumn{2}{c}{\textbf{Active Creators}}\\
  \cmidrule(lr){3-4} \cmidrule(lr){6-7} \cmidrule(lr){9-10}
  & Methods&YouTube & Amazon & &YouTube &Amazon & &YouTube &Amazon  \\
  \midrule
  & w/o Signal & $101.211\pm3.941$ & $29.55\pm0.896$ & & $3.177\pm0.006$ & $2.546\pm0.021$ & & $11.750\pm0.153$ & $9.665\pm0.074$ \\
  \midrule
  \multirow{2}{*}{Naive}
 & MHC & $105.831\pm3.058$ & $57.913\pm1.828$ & & $3.197\pm0.012$ & $2.805\pm0.006$ & & $14.730\pm0.214$ & $11.610\pm0.175$ \\
 & MC & $104.863\pm1.746$ & $\underline{60.123}\pm0.538$ & & $3.186\pm0.018$ & $\underline{2.847}\pm0.008$ & & $13.840\pm0.109$ & $\underline{11.775}\pm0.082$\\
  \midrule
  \multirow{2}{*}{Re-Ranking }
  & P-MMF & $101.599\pm1.263$ & $47.794\pm1.114$& & $3.196\pm0.009$ & $2.697\pm0.011$ & & $15.230\pm0.120$ & $11.025\pm0.097$\\
 & FairRec & $97.171\pm3.396$ & $38.860\pm0.747$ & & $3.182\pm0.011$ & $2.646\pm0.009$ & & $14.245\pm0.153$ & $9.880\pm0.068$\\
  \midrule
  \multirow{1}{*}{Reward Adjust}
  & Reweight & $\underline{110.891}\pm2.505$ & $47.412\pm1.290$& & $\underline{3.223}\pm0.004$ & $2.744\pm0.013$ & & $\underline{15.625}\pm0.127$ & $10.530\pm0.082$\\
  \midrule
  \multirow{1}{*}{Ours}
  & LoRe & $\bm{127.113}\pm2.589$ & $\bm{81.36}\pm0.981$& & $\bm{3.259}\pm0.012$ & $\bm{3.462}\pm0.007$ & & $\bm{16.250}\pm0.148$ & $\bm{14.710}\pm0.164$\\
 \bottomrule
\end{tabular}}

\label{tab:main_table}
\end{table*}



\textbf{Baselines.} We compared our method with several baselines, including the following:
\textbf{w/o Signal}: revealing no information to the creator at all, except for their own creation history. 
Some naive suggestion strategies are as follows. \textbf{Most History Click}: suggesting each creator to produce the item genre with the highest number of user clicks across all genres the creator has ever created;
\textbf{Most Click}: suggesting each creator to produce the item genre with the highest number of user clicks across all genres. In addition to these naive suggestion strategies, we compared our method with approaches that focus on recommendation algorithm design, such as \textbf{P-MMF}~\cite{xu2023pmmf} and \textbf{FairRec}~\cite{guo2023fairrec}. We also included a baseline that influences creators by adjusting their rewards rather than information revelation, namely \textbf{Reweight}~\cite{yao2024user}.

\textbf{Metrics.} Following previous works~\cite{ye2025creagent, yao2024user}, we measure the long-term welfare of RS using the cumulative number of user clicks. We adopt RecAgent~\cite{wang2025user} as the user simulator, in which users decide whether to click on or skip a presented item. The total number of clicks across all users in each round is recorded as the user welfare for that round.
\begin{figure}
\includegraphics[width=1\linewidth]{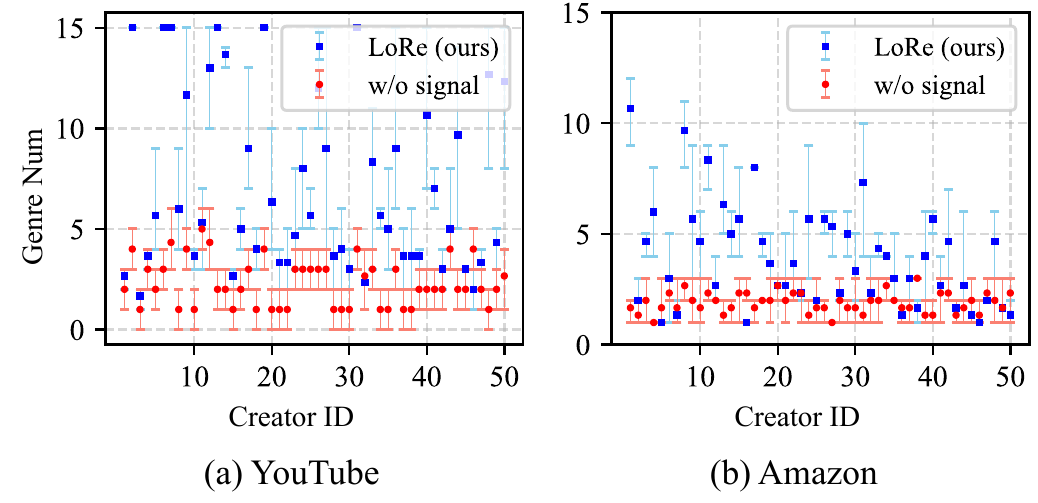}
	\caption{Number of genres created per creator. The blue box denotes the mean number of genres created by each creator under LoRe, while the red box denotes the mean under the w/o Signal baseline.}
	\label{fig:RQ2_2}
\end{figure}
\subsection{Effectiveness of LoRe (RQ1)}
To evaluate the effectiveness of LoRe in enhancing long-term user welfare, we compare its performance against several baselines on two recommendation datasets. In this experimental setting, the number of creators is fixed at 50 and the number of users at 100. The platform employs DIN~\cite{zhou2018deep} as the based recommendation model, and both user and creator agents are powered by LLaMA3-8B~\cite{touvron2023llama}.

Figure~\ref{fig:RQ1_main} illustrates the evolution of user clicks per round as the number of interaction rounds increases, while Table~\ref{tab:main_table} reports the average number of user clicks over $T$ rounds for all methods and datasets. Across both the YouTube and Amazon datasets, LoRe consistently achieves higher user welfare than all baselines. Compared with methods that focus solely on adjusting recommendation algorithms or creator rewards (e.g., PMMF, FairRec, and Reweight), LoRe exhibits substantially better performance, indicating that influencing creators' creation behavior through information revelation is a more effective mechanism for improving long-term user welfare.

Existing baselines also exhibit inconsistent performance across datasets. For example, Reweight outperforms other baselines on the YouTube dataset but performs worse than naive information revelation strategies on the Amazon dataset. This discrepancy is primarily attributable to the more skewed user preference distribution in the Amazon dataset, where a large proportion of users favor a single content genre, while relatively few users prefer other genres. Under such conditions, the performance of existing methods can degrade significantly.

Naive information revelation strategies (Most History Click and Most Click) likewise demonstrate unstable performance: they outperform the w/o Signal baseline on the Amazon dataset but perform comparably on the YouTube dataset. This further indicates that naive strategies lack robustness across different recommendation environments. In contrast, LoRe consistently achieves the highest long-term welfare across all settings, demonstrating that carefully designed information revelation is not only more effective than naive approaches but also more robust across diverse environments.

\subsection{Inherent insight of LoRe (RQ2)}
To better understand the inherent insight of LoRe, we analyze the dynamics of RS from both macro and micro perspectives. At the macro level, we examine the content diversity and the average number of active creators during the interactions among the platform, creators, and users. At the micro level, we investigate how the number of item genres produced by individual creators varies across different methods.

\begin{figure}
  \includegraphics[width=\linewidth]{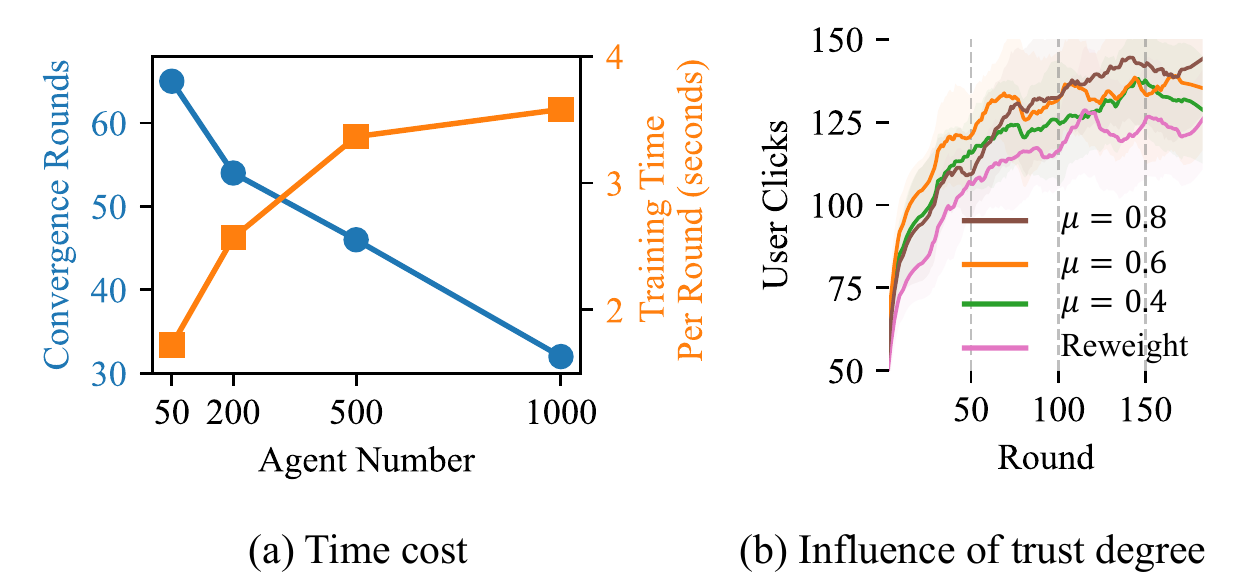}
  \caption{Training time cost and the influence of trust degree. The left figure presents the number of convergence rounds and training time per rounds for LoRe under different agent number. The right figure shows the user clicks of LoRe under different trust degree.}
  \label{fig:cost_and_trust}
\end{figure}

\begin{figure*}
    \includegraphics[width=0.95\linewidth]{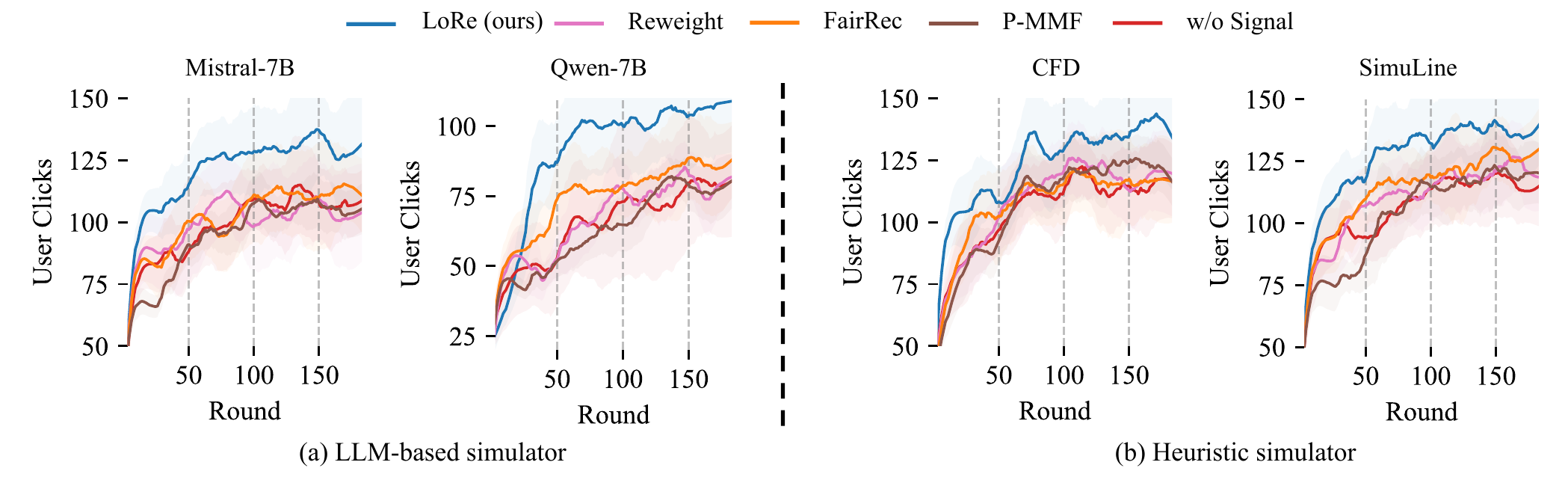}
    \caption{The performance of different methods under various creator simulation models.}
    \label{fig: creator_ablation}
    \vspace{-0.3cm}
\end{figure*}

\textbf{Macro perspective.} To evaluation the content diversity, we employ entropy which is a widely used metric for measuring diversity~\cite{spellerberg2003tribute}. Let $p$ denote the probability density function over genre set $\mathcal{G}$, where $p(g_i)\ (g_i\in\mathcal{G})$ represents the probability of content genre $g_i$. Then, the content diversity is defined as $D(p) = \sum_{g_i\in\mathcal{G}} -p(g_i)\log p(g_i)$. Diversity is an important property of RS, and increasing diversity is often associated with improved user welfare~\cite{castells2021novelty}. Intuitively, higher diversity reflects a more balanced distribution of content genres, whereas lower diversity corresponds to a skewed content distribution in which most content is concentrated in only a few categories. Such extreme content distribution may prevent some users from accessing their preferred content, resulting in user churn and reduced user welfare. Therefore, greater diversity is more conducive to improving users’ long-term welfare. As shown in Table~\ref{tab:main_table}, our method achieves the highest content diversity across different datasets. This demonstrates that our approach enhances the diversity of RS compared with other baselines, which in turn indirectly improves user welfare.

In our simulations, a creator would exit after receiving no clicks for 10 consecutive rounds. We calculated the average number of active creators under different methods across datasets. As shown in Table~\ref{tab:main_table}, our method retains more creators than other baselines. This improvement arises because the platform can influence creators’ behavior by suggesting them creating specific content genre, thereby guiding them to produce content that attracts more clicks.

\textbf{Micro perspective.} We further analyze the variation in the number of content genres produced by each creator. As shown in Figure~\ref{fig:RQ2_2}, the x-axis represents the creator ID, and the y-axis indicates the number of distinct genres created by each individual. Obviously, compared to the baseline w/o Signal, our approach leads to an increase in the number of genres per creator, which directly contributes to the overall improvement in content diversity.

\subsection{Robustness Analysis (RQ3)}\label{sec:creator_ablation}
In this section, we conduct extensive experiments to evaluate the robustness of our method. 

\textbf{Training cost.}
We evaluate the convergence rounds and net training time of our algorithm at different scales to demonstrate its scalability in large-scale simulations. As shown in Figure~\ref{fig:cost_and_trust} (a), our algorithm requires only a few rounds to converge. When the number of agents is 50, it converges in fewer than 70 rounds; when the number of agents increases to 1,000, convergence is achieved in fewer than 40 rounds. In addition, the net training time per round remains within an acceptable range. Even with 1,000 agents, the net training time per round is less than 4 seconds. These results confirm that our method scales effectively to large-scale scenarios.

\textbf{Influence of creator simulation models.}
To evaluate our method's robustness, we conduct experiments under multiple creator simulation models. Specifically, we consider both LLM-based simulation models: CreAgent~\cite{ye2025creagent} (with Mistral-7B and Qwen2.5-7B as backbones), and heuristic-based models: Creator Feature Dynamics (CFD)~\cite{lin2024user,zhan2021towards}, in which creators adjust their creation behavior using user feedback as gradients scaled by a learning rate, and SimuLine~\cite{zhang2023simuLine}, where creators' creations are determined by probabilistic sampling based on the number of clicks in previous steps.

When LLM-based simulation models are used, the platform’s suggestions can be directly incorporated into creators’ prompts, thereby influencing their behavior. However, this mechanism is not applicable to heuristic-based models. To simulate the effect of platform suggestions in such settings, inspired by ~\citet{prasad2023content}, we design a trust-based model. Specifically, each creator $c_i$ is assigned a trust degree $d_i \in [0,1]$, representing the probability that the creator follows the platform’s suggestion upon receiving it. In our experiments, creators’ trust degrees are initialized using a truncated Gaussian distribution to reflect real-world scenarios in which agents typically exhibit moderate trust, while extremely high or low trust levels are uncommon~\cite{teacy2006agent,teacy2012efficient}. Denoted by $\mathcal{N}(\mu, \sigma; 0, 1)$ a truncated Gaussian distribution with mean $\mu$, variance $\sigma$, lower bound $0$, and upper bound $1$. Then, each creator’s trust degree is randomly sampled from distribution $\mathcal{N}(0.5, 1.0; 0, 1)$.

Figure~\ref{fig: creator_ablation} reports the performance comparison between our method and the baselines under different creator simulation models. The results show that, regardless of whether the simulation model is LLM-based or heuristic-based, our approach consistently outperforms all baselines. In contrast, existing recommendation-algorithm-based methods (i.e., P-MMF and FairRec) exhibit inconsistent performance across different creator simulation models. For instance, FairRec performs best among baselines under the Mistral-7B, Qwen2.5-7B, and SimuLine simulation models but is outperformed by P-MMF under the CFD model. Moreover, Reweight, which influences creator behavior by adjusting their rewards, performs relatively poorly. This is because it is designed for a specific creator decision model and thus may fail to generalize well when the decision model changes.

\textbf{Influence of creators' trust degree.}
When heuristic-based models are used to simulate creators' behavior, we assume that each creator has a degree of trust on the platform. In real-world RS, creators naturally differ in their trust. Highly trusting creators are more likely to follow platform suggestions, whereas creators with low trust are less susceptible to such influence. Moreover, creators' trust may evolve over time as a function of the feedback they receive\cite{prasad2023content}. To evaluate the effectiveness of our method under different levels of creator trust, we conduct experiments where creators' initial trust degrees are sampled from distributions with different expectations.
Specifically, we conduct experiments with $\mu \in \{0.4, 0.6, 0.8\}$ and $\sigma=1.0$. Moreover, we assume that creators’ trust degrees change dynamically over time. Let $d_i^{t}$ denote creator $i$’s trust degree in round $t$ and $r_i^{t}$ represent the number of clicks received by creator $i$ in round $t$. Then, creator $i$’s trust in round $t+1$ is updated as $d_i^{t+1}=d_i^t+\frac{r_i^{t+1} - r_i^{t}}{r_i^{t}}$, if they follow the platform’s suggestion. This means that a creator’s trust increases if following the platform’s suggestion yields more clicks, and decreases otherwise.

Figure~\ref{fig:cost_and_trust} (b) shows the user clicks under trust degree distributions with different expectations on the YouTube dataset. We compare our method with Reweight, since it performs best among the baselines on YouTube. The results show that although our method's performance fluctuates slightly across different trust distributions, it consistently outperforms Reweight, demonstrating robustness to varying levels of creator trust. Furthermore, while performance of our method decreases slightly as the expected trust degree decreases, the decline is marginal. Overall, these findings underscore the importance of maintaining platform credibility by providing consistently beneficial suggestions to creators.

\begin{figure}
    \centering
    \includegraphics[width=0.95\linewidth]{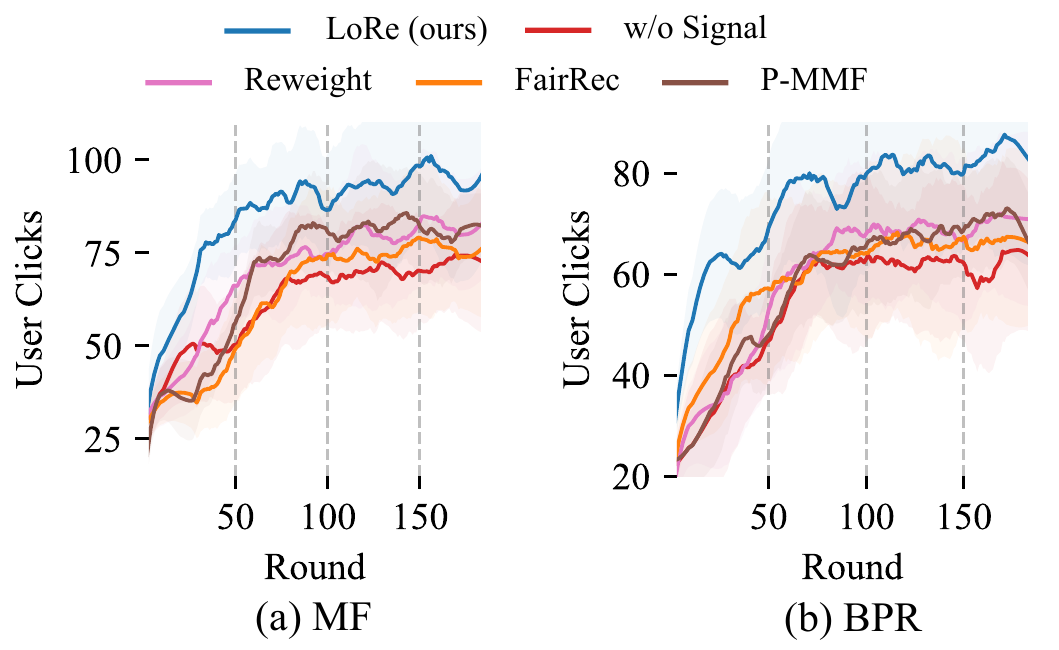}
    \caption{Comparison of user welfare with our method and baselines under the RS environment with two different recommendation models, i.e, MF and BPR, respectively.}
    \label{fig:rec_model}
\end{figure}
\textbf{Influence of different recommendation algorithms.}
In real-world RS environments, platforms employ different recommendation algorithms to select content from the item pool for users.
To verify the effectiveness of our method under various recommendation models, we conduct experiments using different recommendation algorithms as the platform's recommendation policy.
Specifically, we select two of the most widely used recommendation models: Matrix Factorization (MF)~\cite{koren2009mf} and Bayesian Personalized Ranking (BPR)~\cite{rendle2012bpr}.
As illustrated in Figure~\ref{fig:rec_model}, our method consistently achieves higher user welfare than the baselines across different recommendation algorithms.
This result demonstrates that the effectiveness of LoRe is not tied to a specific recommendation model, but instead generalizes well across different algorithmic.

\begin{table}
  \caption{The average user clicks achieved through combining LoRe with FairRec or P-MMF on different datasets.}
  \label{tab:RQ3_4}
  \begin{tabular}{lcc}
    \toprule
     Method & YouTube & Amazon\\
    \midrule
    P-MMF & $108.569\pm 1.213$ &  $51.275\pm 0.984$\\
    FairRec & $113.279\pm 1.146$ & $46.187\pm 0.813$\\
    LoRe (ours) & $137.495\pm 2.523$ & $95.150\pm 0.687$  \\
    LoRe + P-MMF & $\underline{141.226}\pm 2.266$ & $\bm{101.997}\pm 1.084$\\
    LoRe + FairRec & $\bm{146.009}\pm 1.742$ & $\underline{98.618}\pm 0.994$  \\
  \bottomrule
\end{tabular}
\end{table}
\textbf{Incorporated with existing methods.}
We integrate our method with existing approaches that focus on recommendation algorithm design, including P-MMF and FairRec. We exclude Reweight from this combination, as both LoRe and Reweight aim to enhance user welfare by influencing creators' creation behavior. Table~\ref{tab:RQ3_4} presents the average number of user clicks achieved by combining our method with these approaches on the YouTube and Amazon datasets. The results show that integrating LoRe with these methods further improves performance across both datasets. In particular, LoRe + FairRec achieves higher user welfare on the YouTube dataset, whereas LoRe + P-MMF performs better on the Amazon dataset. These findings demonstrate that LoRe is compatible with existing recommendation algorithm and highlight its scalability.
\section{Social Impacts}

Our method effectively enhances long-term user welfare in recommendation systems. Beyond improving user satisfaction and engagement over time, it also contributes to the development of a healthier and more sustainable ecosystem for content creation and consumption. A potential concern is that platform might misuse such suggestion strategy to manipulate creators' behavior, benefiting some creators at the expense of others. However, the proposed suggestion strategy does not compel creators' actions, that is creators retain full autonomy over the content they create. Moreover, if the platform were to provide suggestions that systematically harm creators' earnings, it would erode the platform’s credibility, reduce creators’ trust, and ultimately diminish user welfare. Thus, platforms lack incentives to issue harmful or manipulative suggestions.

\section{Conclusions}

In this paper, we demonstrate that influencing creators' creation behavior could be a better way to improve long-term user welfare compared to methods relying on recommendation algorithm design. Inspired by the idea of cheap talk, we model the platform as the sender and creators as the receivers. To address the challenges of applying cheap talk to RS, we propose a novel method, LoRe, which formulates the dynamic interaction between the platform and creators as a MDP. LoRe employs a reinforcement learning algorithm to optimize the platform’s suggestion strategy. Extensive experiments on two real-world datasets demonstrate the effectiveness of our approach and show that it can be seamlessly integrated into the existing RS ecosystem, as well as combined with other methods that focus on recommendation algorithm design.


\bibliographystyle{ACM-Reference-Format}
\bibliography{sample-base}


\end{document}